# Machine Learning Potential-Driven Molecular Dynamics Simulations of Dehydrogenation in Pristine and Doped $MgH_2$

Bo Han[a,b], Jianchuan Wang[a,*], Rui Zhang[b,c], Martin Matas[b,d], Elias Vigl[e], Jing Tang[b], Yong Du[a], Christian Weiss[e], David Holec[b,*]

[a] State Key Laboratory of Powder Metallurgy, Central South University, 410083, Changsha, China

[b] Department of Materials Science, Montanuniversität Leoben, 8700, Leoben, Austria

[c] Department of Materials Science, Key Laboratory of Automobile Materials, MOE and State Key Laboratory of High Pressure and Superhard Materials, International Center of Future Science, Jilin University, Changchun 130012, China

[d] Department of Physics and NTIS – European Centre of Excellence, University of West Bohemia in Pilsen, Univerzitní 8, 301 00 Pilsen, Czech Republic

[e] Chair of Process Technology and Industrial Environmental Protection, Montanuniversität Leoben, 8700, Leoben, Austria

* Corresponding author. Email: jcw728@126.com, david.holec@unileoben.ac.at

**Abstract**

Machine learning potential-driven molecular dynamics simulations (ML-MD) were employed to provide atomistic insights into the dehydrogenation kinetics of pristine and doped $MgH_2$. Through systematic investigation of distinct surface orientations, $MgH_2$(100) surface was identified as the most active low-index surface for hydrogen release. For pristine $MgH_2$, our simulations revealed a novel $H_2$ formation mechanism characterized by $H_2$ generation in the subsurface region followed by diffusion to the surface for desorption, highlighting the critical role of subsurface processes beyond conventional surface-driven pathways. Comprehensive screening of 22 doping elements identified Ni as the most effective dopant. Among several descriptors, machine learning analysis identified the time-coupled Miedema electron density as the critical descriptor, underscoring the role of electronic properties. Consequently, a volcano-shaped relationship was uncovered between the intrinsic Miedema electron density ($n_{ws}$) and total hydrogen release (optimal window: $4.0 < n_{ws} < 5.4 \times 10^{-2}$ e/bohr$^3$). Dopants within this range serve a dual function: acting as thermodynamic sinks for H attraction while maintaining a balanced interaction strength to facilitate H–H coupling and $H_2$ release. This atomistic-level validation provides strong theoretical support for the experimentally observed “hydrogen pump” effect of catalytic phases. The present study demonstrates the strong capability of ML-MD in navigating through complex catalytic mechanisms and establishing quantitative property–activity relationships, providing a robust framework for rational design of high-performance catalysts for $MgH_2$ and other hydrogen storage materials.

## 1. Introduction

Hydrogen gas, as a clean, efficient, and sustainable energy carrier, has garnered significant attention in the global transition towards low-carbon energy systems. Among various hydrogen storage technologies, solid-state hydrogen storage materials—particularly metal hydrides—have attracted considerable interest due to their high volumetric hydrogen density, safety, and reversibility. Magnesium hydride ($MgH_2$) stands out as one of the most promising candidates, offering a high gravimetric hydrogen capacity of 7.6 wt%, abundant raw material reserves, and low cost [1, 2]. However, its practical application is severely hindered by high dehydrogenation enthalpy (~76 kJ/mol $H_2$) and sluggish absorption/desorption kinetics [3-5]. The mentioned drawbacks are primarily attributed to the strong ionic bonds within the $MgH_2$ lattice and the high activation energy barrier for hydrogen diffusion [6].

To overcome these challenges, reducing the particle size of $MgH_2$ to the nanoscale and doping with catalysts have proven to be efficient approaches [7-9]. Nanostructuring not only shortens the diffusion distance for hydrogen atoms but also significantly increases the specific surface area, thereby exposing a greater number of active sites for dehydrogenation. Catalyst doping, on the other hand, facilitates hydrogen bond breaking and provides additional catalytic active sites, further lowering the dehydrogenation barrier. Although experimental approaches such as ball milling can efficiently prepare nanosized $MgH_2$, it remains challenging to experimentally observe the atomic-scale dehydrogenation mechanisms. In particular, understanding how specific surface orientations and surface dopants influence hydrogen desorption dynamics is difficult to achieve through experimental characterization alone. This highlights the urgent need for theoretical approaches to complement experiments and provide fundamental insights into the surface-mediated dehydrogenation of $MgH_2$.

In recent years, molecular dynamics (MD) simulation has emerged as a powerful tool for exploring atomic-scale dynamics and reaction mechanisms. However, conventional MD approaches face significant bottlenecks in this context. Classical

MD simulations rely heavily on the accuracy and availability of interatomic potentials. While potentials exist for the binary Mg-H system [10, 11], extending them to multi-component Mg-H-M systems remains a considerable challenge. This limitation is particularly pronounced in doping studies, where developing system-specific potentials for each dopant combination is not only labor-intensive but also prone to inaccuracies when capturing the chemical complexity of bond breaking and formation. In contrast, *ab initio* MD (AIMD) avoids the need for empirical potentials by directly solving the electronic structure at each MD step, thereby ensuring quantum-mechanical accuracy. However, this comes at a prohibitive computational cost, restricting simulations to small systems and short timescales.

To address these limitations, machine learning interatomic potentials (MLIPs) have emerged as a revolutionary approach, combining the quantum-mechanical accuracy of density functional theory (DFT) with the computational efficiency typical of classical force fields [12]. By training on vast DFT datasets of energies and forces, MLIPs can learn the underlying potential energy surface, enabling accurate simulations of large systems over extended timescales [12]. This is particularly valuable for investigating complex chemical scenarios, such as the interaction of hydrogen with various surface facets and dopants, which are computationally prohibitive for traditional methods.

Among emerging MLIPs, the 'PreFerred Potential' (PFP) [13, 14] has emerged as a universal framework trained on diverse chemical systems across the periodic table, incorporating configurations relevant for ion transport and surface phenomena. It offers unprecedented transferability without the need for system-specific retraining, a capability that has recently enabled various complex computational studies [15-21]. Leveraging this advantage, we employed PFP to perform large-scale machine learning potential-driven molecular dynamics (ML-MD) simulations to investigate the dehydrogenation of $MgH_2$. These simulations extended up to 50 ps with systems containing approximately 2700 atoms—time and length scales well beyond the reach of AIMD. We systematically investigated the dehydrogenation across distinct low-index $MgH_2$ surfaces and comprehensively screened a wide range of dopants. By integrating dynamic structural analysis with electronic property correlations, we aim to elucidate the complete reaction mechanism from subsurface nucleation to surface desorption and thereby establish a quantitative property-activity relationship. This work provides a robust framework for the rational design of high-performance catalysts for hydrogen storage materials.

## 2. Methods

### 2.1 Structural models

Four low-index $MgH_2$ surfaces were selected for investigation: (110), (100), (101), and (001). These surfaces were constructed by cleaving the bulk $MgH_2$ crystal (space group $P4_2/mnm$) along the respective Miller indices. The bulk structure utilized optimized lattice constants from our previous work ($a = b = 4.487$ Å and $c = 2.994$ Å) [22], which are in agreement with the experimental values ($a = b = 4.501$ Å and $c = 3.010$ Å) [23]. For each orientation, supercell slab models with a thickness of approximately 37 Å were constructed to ensure that the central atomic layers exhibit bulk-like properties. The number of Mg atomic planes (hereafter referred to as "Mg layers") was systematically varied to account for the distinct atomic packing and layer spacing along different crystallographic directions. Specifically, 12, 17, 15, and 25 Mg layers were employed for the (110), (100), (101), and (001) surfaces, respectively. A vacuum space of 100 Å perpendicular to the slab surface was introduced to eliminate periodic interactions in the surface normal direction. The optimized surface supercell structures are presented in Fig. 1, and the corresponding structural parameters along with calculated surface energies are listed in Table S1.

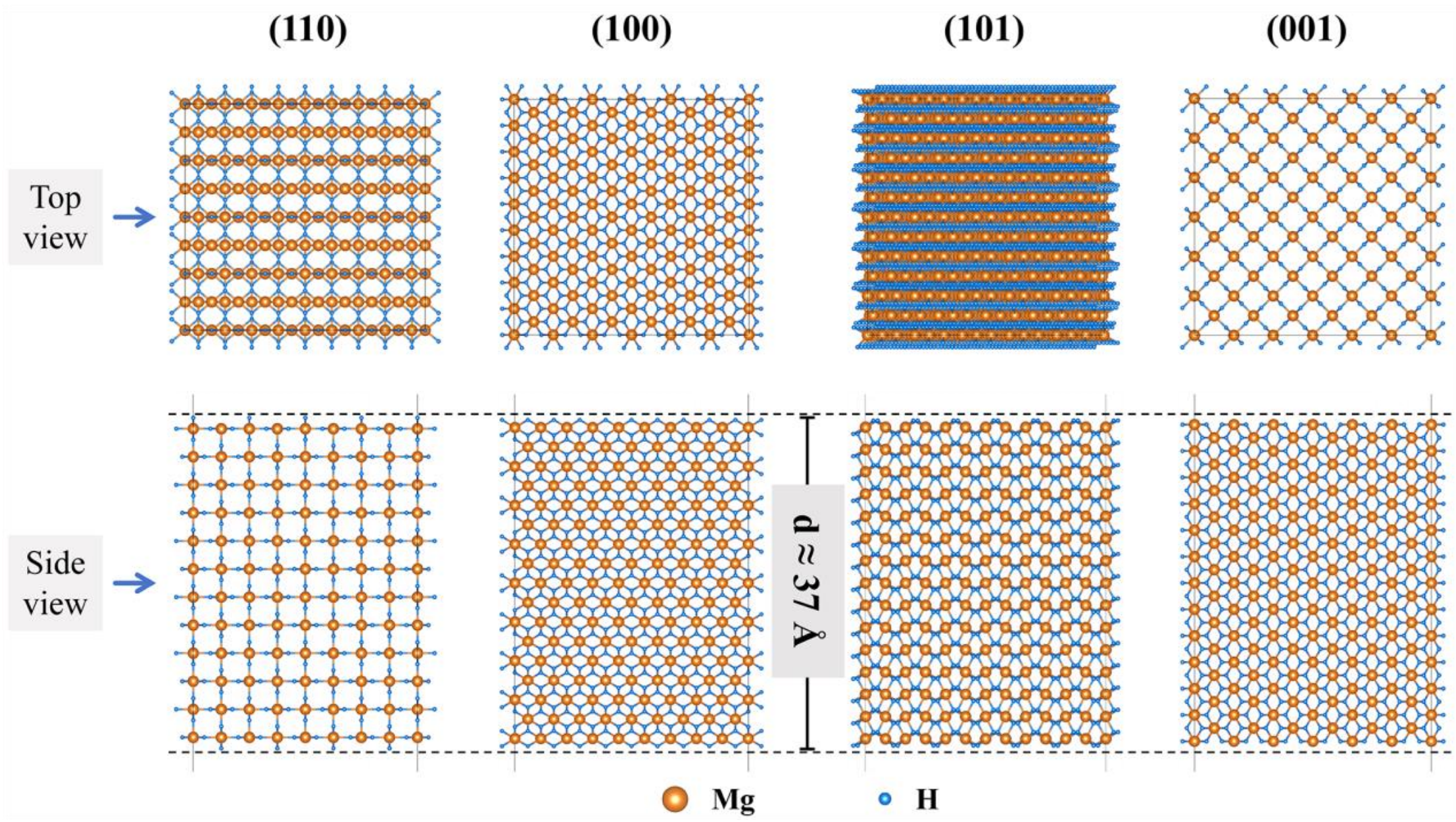

Fig. 1. Atomic structures of $MgH_2$ (110), (100), (101), and (001) surfaces.

## 2.2 Calculation details

### 2.2.1 Matlantis potential

In this work, we employed the universal neural network potential known as the ‘PreFerred Potential’ (PFP, version 7.0.0), which incorporates van der Waals (vdW) corrections. This potential was developed by Preferred Computational Chemistry (PFCC) and is accessible through the cloud-based simulation platform Matlantis$^{TM}$ [13, 14]. PFP is capable of describing interatomic interactions across 96 elements of the periodic table, making it highly versatile for diverse material systems. The PFP architecture is based on the TeaNet framework, augmented with a More-style two-body potential to handle short-range repulsion [14]. The training and validation datasets for PFP were generated by first-principles calculations based on DFT using the Vienna *Ab initio* Simulation Package (VASP) [24, 25], with the training set comprising 59 million structures. The electron-ion interaction was described via the projector augmented wave (PAW) method with a kinetic energy cutoff of 520 eV [26, 27]. The exchange-correlation effects were treated by the Perdew-Burke-Ernzerhof (PBE) functional [28]. Gaussian smearing with a width of 0.05 eV was applied, and the *k*-point density was set to 1000 *k*-points per reciprocal atom [13]. The training procedure utilized the system energy, atomic forces, and atomic charges as target properties. Further details regarding the potential generation can be found in the original references [13, 14].

### 2.2.2 Molecular dynamics simulations

To capture $H_2$ desorption events within the limited timescale of molecular dynamics (MD) simulations and to obtain statistically significant data, the simulation temperature was set to 1000 K for both the undoped and doped systems. Although this exceeds the experimental desorption temperature of $MgH_2$ (~683 K) [29], such an elevated temperature effectively overcomes kinetic barriers and significantly accelerates the dehydrogenation process, enabling sufficient sampling of H2 desorption events for statistical analysis. In addition, for the undoped system, a separate MD simulation at a lower temperature of 800 K, only slightly above the experimental desorption temperature, was performed to observe the dehydrogenation mechanism with greater fidelity and mechanistic clarity, avoiding the risk that an excessively accelerated dynamics at 1000 K would obscure the fine structural details of the process.

To mitigate initial fluctuations in energy and temperature, the system was initialized at a temperature 200 K above the respective target temperature (1200 K for the 1000 K runs, 1000 K for the 800 K run). The simulations were carried out with a time step of 0.5 fs in the canonical (NVT) ensemble using the Nosé-Hoover thermostat [30]. Each trajectory consisted of 100,000 timesteps, corresponding to a total simulation time of 50 ps.

To maintain a low hydrogen partial pressure consistent with experimental conditions, $H_2$ molecules were removed from the system once they diffused beyond a vertical distance of 30 Å from the surfaces. This operation prevents the subsequent $H_2$ desorption from being hindered by the excessive $H_2$ gas outside pressure due to the limited MD simulation volume. To demonstrate the necessity and rationality of this operation, we calculated the pressure change in the original vacuum layer in the presence of different numbers of $H_2$ molecules, as shown in Fig. S1. In the presence of five $H_2$ molecules, the pressure reaches approximately 9 atm.

The formation of an $H_2$ molecule was defined when the interatomic distance between two hydrogen (H) atoms fell below 0.85 Å. While the equilibrium H-H bond length is approximately 0.74 Å, this threshold was intentionally set at 0.85 Å to account for the increased bond vibration amplitude due to thermal fluctuations at high temperatures. To minimize statistical errors, five independent MD simulations were conducted for each system, and the results were averaged. All calculations and post-processing analyses were conducted within the Matlantis platform (Jupyter/Python environment), utilizing the Atomic Simulation Environment (ASE) toolkit [31].

### 2.3 Machine learning-based descriptor analysis

To elucidate the intricate correlations between dopant properties and dehydrogenation kinetics, we developed an advanced machine learning framework employing the eXtreme Gradient Boosting (XGBoost) method [32] coupled with SHapley Additive exPlanations (SHAP) analysis [33]. Our dataset comprised 11,000 data points derived from time-resolved MD simulations across 22 distinct dopant systems. The cumulative $H_2$ release after 50 ps was used as the primary target variable for kinetic assessment.

Machine learning analyses were performed using the Scikit-learn and XGBoost libraries within a Python 3.11 environment [34]. An initial set of 11 elemental descriptors was extracted from the Mendeleev library, encompassing three key categories: (1) fundamental structural parameters (e.g., atomic radius, covalent radius), (2) critical electronic properties (e.g., Miedema electron density, electron affinity), and (3) essential thermodynamic attributes (e.g., melting point, specific heat capacity). To precisely capture the dynamic evolution of dehydrogenation, these descriptors were multiplied by normalized time to engineer a suite of time-coupled descriptors. Specifically, to prevent interference from static attributes, the model input was restricted exclusively to these time-coupled descriptors. For simplicity, these descriptors are denoted by symbols as detailed in Table S2. All descriptors were rigorously standardized prior to model training to ensure optimal performance.

The XGBoost model was trained on 80% of the preprocessed dataset (8,800

samples), with hyperparameters optimized via 10-fold cross-validation. The optimal configuration was determined as: max_depth = 6, learning_rate = 0.2, and n_estimators = 100. This model achieved an $R^2$ score of 0.998 on the test dataset, demonstrating high predictive accuracy and strong generalization capability. To quantitatively assess the interpretability and relative influence of each descriptor, descriptor importance was evaluated using mean absolute SHAP values, which were computed via the TreeExplainer algorithm on the training dataset.

### 2.4 First-principles calculations

First-principles calculations were carried out using VASP based on DFT [24, 25]. The electron-ion interaction was described via the PAW method [26, 27], while the exchange-correlation effects were treated by the PBE functional of the generalized gradient approximation (GGA) [28]. The reciprocal space of the $MgH_2$ slab supercell ($a$ = 8.977 Å, $b$ = 8.983 Å, and $c$ = 24.855 Å) was sampled using a 4×4×1 Monkhorst-Pack grid [35], which was found to be sufficient to converge the free energy to within 0.1 meV/atom. The cut-off energy was set to be 375 eV for plane-wave expansions. The absolute total energy changes and the average force per atom were converged to $1.0\times10^{-5}$ eV and 0.02 eV/Å for electron self-convergence calculation and structural relaxation, respectively. All structural optimizations were performed using the conjugate gradient method.

### 2.5 Experimental details

To complement the theoretical investigation, the dehydrogenation characteristics for pure and Ni-containing magnesium hydride powders were examined to illustrate the influence of transition metals in technical storage materials. Both materials were produced from gas-atomized metallic precursors (pure Mg as well as MgNi15 alloy, Metalpine GmbH) with comparable particle size distributions (mass median diameters and cumulative 90 m.% diameters: $D_{50}$ = 28.1 µm, $D_{90}$ = 47.9 µm for Mg powders and $D_{50}$ = 24.9 µm, $D_{90}$ = 42.5 µm for the MgNi15 alloy). The precursors were loaded with hydrogen (99.999 %, Linde) at a temperature of 623 ± 10 K for 24 h at a pressure > 2.0 MPa (initial pressure: 3.5 ± 0.1 MPa). Simultaneous thermal analysis (STA) was performed with a Netzsch STA 449C in open $Al_2O_3$ crucibles under 60 mL/min Ar (99.999%, Linde). For the estimation of the activation energies of desorption, the Kissinger method [36] was applied using heating rates $\beta$ of 2.5, 5, 10 and 20 K/min. For the pure $MgH_2$, peak maximum temperatures $T_p$ were determined directly from the differential scanning calorimetry (DSC) signal. In the case of the Ni-containing samples, the main DSC peak consisted of two overlapping peaks arising from two distinct hydrogen-release processes. These peaks were therefore deconvoluted into two components by fitting a model consisting of two Pearson IV functions (see Fig. S5-S7). Note that the 5 K/min data for the Ni-containing sample was excluded from the final Kissinger analysis due to severe peak overlap, as will be detailed later. The resolved peak maxima from the remaining heating rates were then used to obtain two individual estimates for the desorption activation energies.

## 3. Results and discussion

### 3.1 Surface orientation dependence

To assess the impact of surface orientation on the dehydrogenation kinetics of $MgH_2$, we quantified the cumulative $H_2$ release across four distinct low-index facets. As illustrated in Fig. 2, the dehydrogenation performance exhibits notable anisotropy, with total hydrogen release following the order: (100) > (001) > (101) > (110). The superior activity of the (100) surface can be attributed to its relatively open structure with larger interstitial voids compared to the (101) and (110) surfaces. This open topology minimizes steric hindrance during $H_2$ desorption, facilitating a more efficient release process. Although the (001) surface exhibits an even more open structure than (100), the coplanar arrangement of its H and Mg atoms restricts H-atom mobility, partially offsetting this structural advantage. From a thermodynamic perspective, surface energy is a crucial determinant of chemical reactivity, as surfaces with higher energies are generally less stable and thus more chemically reactive. The (110) surface, exhibiting the lowest surface energy of 0.34 J/m$^2$, demonstrates the least activity due to its stability and reduced propensity for bond breaking. The (101) surface, with a higher surface energy of 0.39 J/m$^2$, shows only moderate dehydrogenation kinetics due to its dense structural characteristics. Although the (001) surface has the highest surface energy at 0.50 J/m$^2$, its activity is limited by the aforementioned structural constraints, which hinder the desorption process. Consequently, after 50 ps, its apparent activity is slightly lower than that of the (100) surface, despite the latter having a significantly lower surface energy of 0.38 J/m$^2$. Overall, the (100) and (001) surfaces appear similarly active, at least within the restricted monitoring time of the simulation. Therefore, to rationalize the observed anisotropic dehydrogenation kinetics, it is necessary to consider both the structural characteristics and the thermodynamic stability of the different facets. Based on this initial test, we focus our further analysis on the active (100) surface, which exhibits the maximum accumulated hydrogen release.

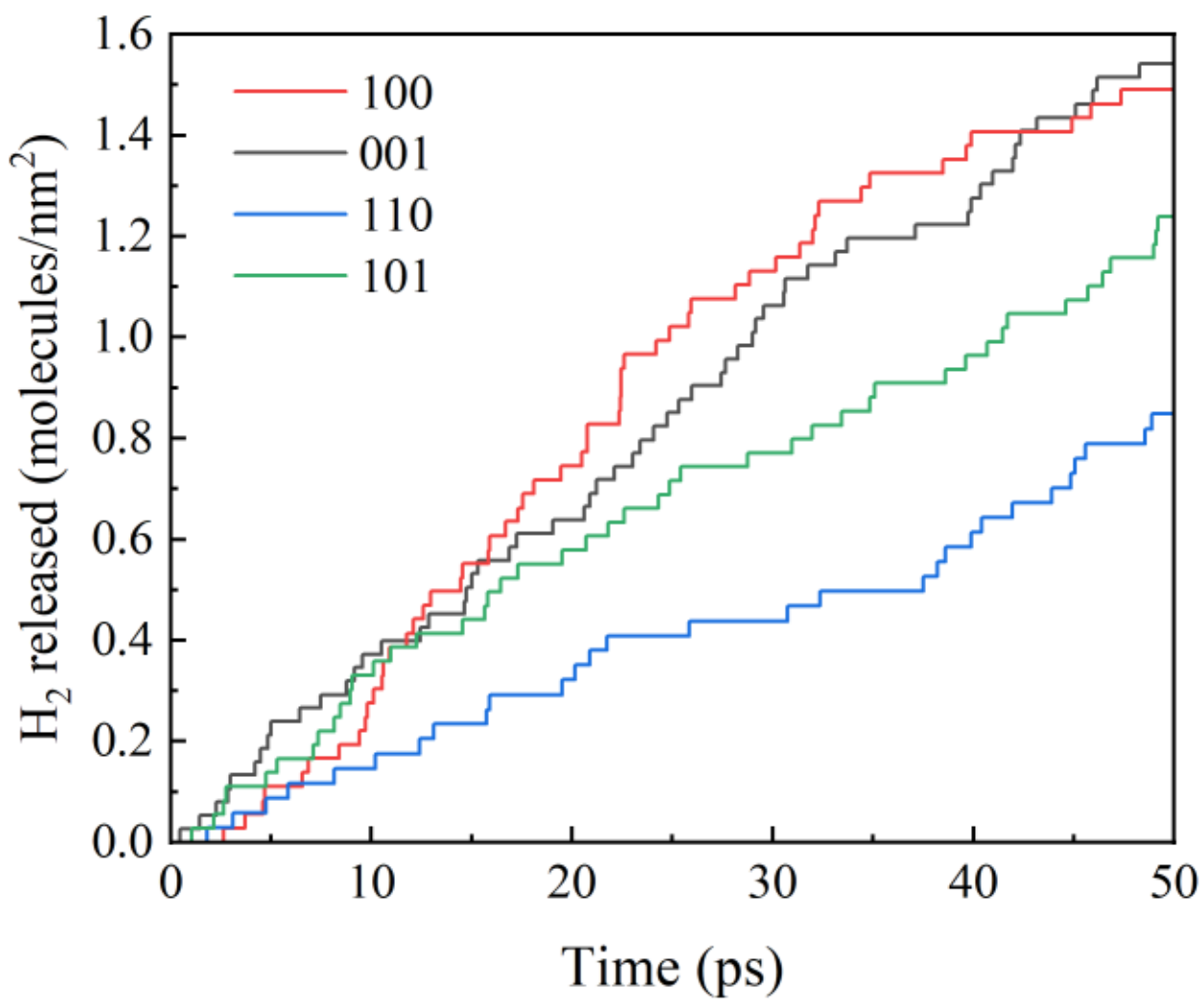

Fig. 2. Dehydrogenation curves of $MgH_2$(100), (001), (101), and (110) surfaces.

### 3.2 Dehydrogenation pathway

Elucidating the atomic-scale pathway of hydrogen release is the first step in understanding dehydrogenation kinetics. To observe the hydrogen desorption process, MD simulations of the pristine $MgH_2$(100) surface were performed at 800 K. This temperature, which is higher than the experimental desorption temperature for $MgH_2$ (~683 K) [29], was chosen to ensure observable events within the accessible simulation timescale.

Contrary to the conventional view that H atoms diffuse to the outermost surface layer to recombine to form $H_2$ and desorb, our MD simulation reveals a distinct subsurface recombination mechanism, as schematically visualized in Fig. 3. Following its formation, an $H_2$ molecule diffuses through the lattice and eventually desorbs into vacuum. This finding aligns with Morrison's theoretical study [37], in which all eight $H_2$ molecules that desorbed during their MD simulations of $MgH_2$(110) surface were not generated on the surface. Rather, they were formed in the subsurface layer and subsequently diffused to the surface. The consistency between our results for the (100) surface and their findings for the (110) surface indicates that subsurface recombination represents a crucial general characteristic of $MgH_2$ dehydrogenation, regardless of surface orientation.

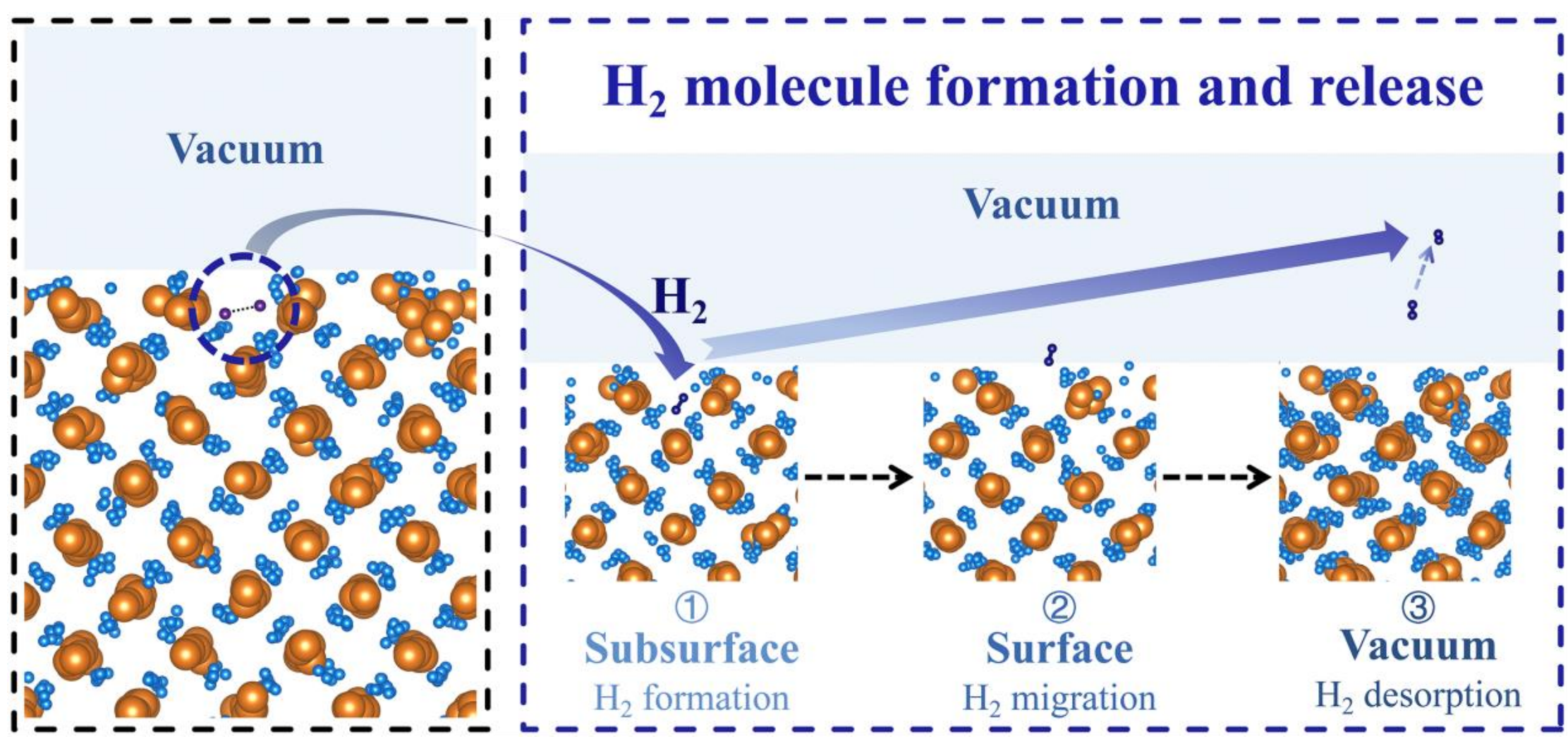

Fig. 3. Schematic illustration of an atomic-scale pathway of hydrogen release from $MgH_2$(100) surface revealed by MD snapshots from the simulation at 800 K. Orange and blue spheres represent Mg and H atoms, respectively. The sequential desorption process shows: (1) $H_2$ formation at the subsurface, (2) migration to the surface, and (3) desorption into vacuum. The formed $H_2$ molecule is highlighted in dark blue with H-H bonds visible. Note: The apparent clustering of H atoms is due solely to the perpendicular view into the lattice interior.

To further validate the subsurface recombination mechanism, we performed first-principles calculations to calculate the desorption energy ($E_{\mathrm{d}}$) of an $H_2$ molecule for $MgH_2$(100) surface using the following equation:

$$E_{\mathrm{d}} = E_{\mathrm{MgH_2}}^{\mathrm{H_2}} - E_{\mathrm{MgH_2}} \quad (1)$$

where $E_{\mathrm{MgH_2}}$ represents the total energy of the pristine $MgH_2$(100) surface supercell,

and $E_{\mathrm{MgH_2}}^{\mathrm{H_2}}$ represents the total energy of the same supercell after an $H_2$ molecule has recombined and been relocated to the vacuum region far from the surface. Both configurations contain the same number of atoms; the $H_2$ molecule is not removed from the system but displaced into the vacuum region of the identical supercell. Therefore, no separate gas-phase $H_2$ energy term is required, and the desorption energy is obtained directly from the difference between the two total energies.

We constructed a p(2×3) $MgH_2$(100) surface supercell with 14 atomic layers containing 84 atoms, as shown in Fig. S2. In addition, a vacuum layer of 15 Å was added along the $Z$-direction to relieve the periodic interaction. The lattice constants of this surface model are $a$ = 8.977 Å, $b$ = 8.983 Å and $c$ = 24.855 Å. The bottom two atomic layers were fixed to mimic bulk condition while other atomic layers were allowed to relax. We selected H atoms located in the top three H atomic layers, denoted as H1, H2 and H3, respectively. The desorption energies of $H_2$ molecule were then calculated by considering different combinations of these H atoms to be removed and recombine into an $H_2$ molecule at the surface. Specifically, we first examined the recombination of two H atoms within the same layer (i.e. H1-H1, H2-H2, H3-H3) and then between adjacent layers (e.g. H2-H3). The calculation results are listed in Table 1. As shown in Table 1, the $E_{\mathrm{d}}$ values for H2-H2 and H3-H3 are 2.13 eV and 2.44 eV, respectively, which are higher than that of H1-H1 (1.96 eV). In contrast, the H2-H3 exhibits the lowest desorption energy among all cases, with a value of 1.59 eV, which is lower by 0.36 eV than that of H1-H1. This suggests that the recombination of H atoms from the second and third subsurface H layers (H2 and H3) is thermodynamically more favorable for $H_2$ desorption compared to combinations within the same inner layer or even the topmost H layer (H1). The lower desorption energy for H2-H3 indicates that H atoms in subsurface layers can more easily recombine and desorb as $H_2$ molecule. This finding supports the subsurface recombination mechanism observed in the ML-MD simulations, and corroborates the reliability of the ML-MD simulation method.

Table 1 Desorption energies ($E_{\mathrm{d}}$) of $H_2$ molecule in $MgH_2$(100) surface

| | Combination of H atoms | | | |
|---|---|---|---|---|
| | H1-H1 | H2-H2 | H3-H3 | H2-H3 |
| $E_d$ (eV) | 1.96 | 2.13 | 2.44 | 1.59 |

This phenomenon, wherein $H_2$ preferentially forms in the subsurface, is primarily attributed to its unique local atomic environment. While hydrogen atoms generally exhibit facile diffusion on pristine surfaces, the outermost surface atoms in actual systems tend to be under-coordinated. This results in strong, localized bonding that effectively traps hydrogen atoms in deep potential wells. Consequently, even if individual atomic hopping events occasionally occur, the effective mobility and collision probability required for recombination are severely restricted. In contrast, the subsurface region provides a unique intermediate environment: the lattice constraints here are sufficiently relaxed to facilitate rapid hydrogen migration without imposing strong trapping potentials. Furthermore, the high local hydrogen density in this region significantly increases the probability of H-H collisions and recombination, further promoting $H_2$ formation. On the other hand, although the deep bulk also possesses a high hydrogen density, $H_2$ recombination in this region is both thermodynamically and kinetically disfavored. Under the rigid three-dimensional confinement of the bulk lattice, the formation of a bulky $H_2$ molecule would induce severe localized lattice distortion, resulting in a substantial strain energy penalty. Conversely, due to its proximity to free surface, the subsurface provides the necessary structural flexibility to effectively accommodate the outward lattice strain released during $H_2$ formation. In summary, the subsurface region achieves an optimal balance among atomic mobility, high recombination probability due to strong trapping, and structural flexibility, making it the ideal zone for $H_2$ formation.

Subsurface recombination, while mechanistically preferred over surface recombination, still faces substantial kinetic barriers arising from strong Mg-H bonding and limited hydrogen mobility within the lattice. Moreover, even after formation, the $H_2$ molecule must diffuse through the overlying Mg-H layers before desorption, posing a second kinetic bottleneck. These insights establish that accelerating $MgH_2$ dehydrogenation requires addressing two fundamental challenges: (1) reducing the barrier for H-H recombination, and (2) facilitating the transport of formed $H_2$ molecules to the surface. The following section explores the potential of doping to overcome these intrinsic limitations.

### 3.3 Dopant screening

Having established the intrinsic dehydrogenation pathway of $MgH_2$, we next investigated whether elemental doping can overcome the kinetic limitations identified in Section 3.2. Given the superior dehydrogenation kinetics of the $MgH_2$(100) surface compared to other low-index surfaces, it was selected as the host substrate for catalyst screening. The selected doping elements are shown in Fig. 4(a). Doped models were constructed by randomly substituting three Mg atoms with doping elements in both the top and bottom surface layers, as illustrated in Fig. 4(b), while the rest of the bulk remains pristine $MgH_2$. The atomic doping concentration relative to the number of Mg atoms in the substituted surface layers is 5.56 at.%. During the subsequent MD simulations, these dopants tend to diffuse inwards rather than be confined to the outermost surfaces. Taking Ni as an example, Fig. 5(a) shows its increasing depth over time, and Fig. 5(b) confirms that all doping elements exhibit inward diffusion.

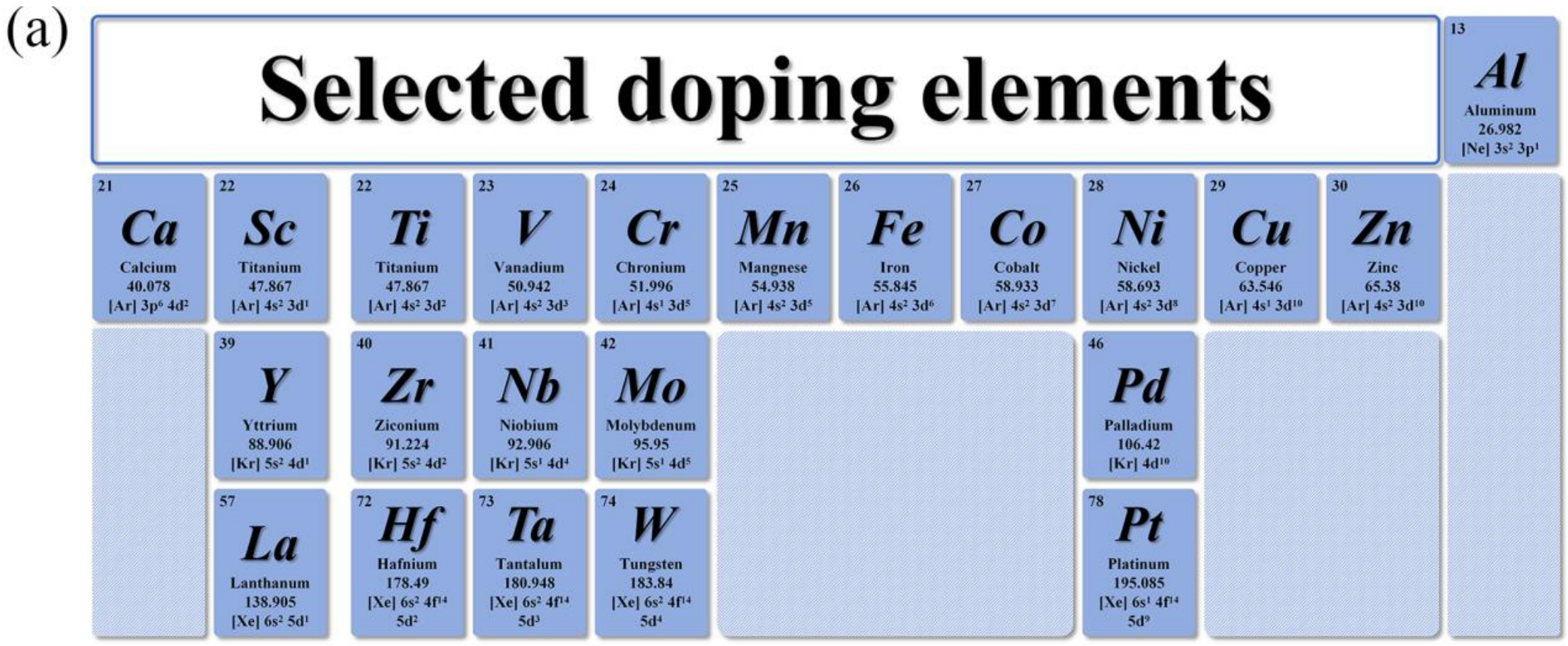

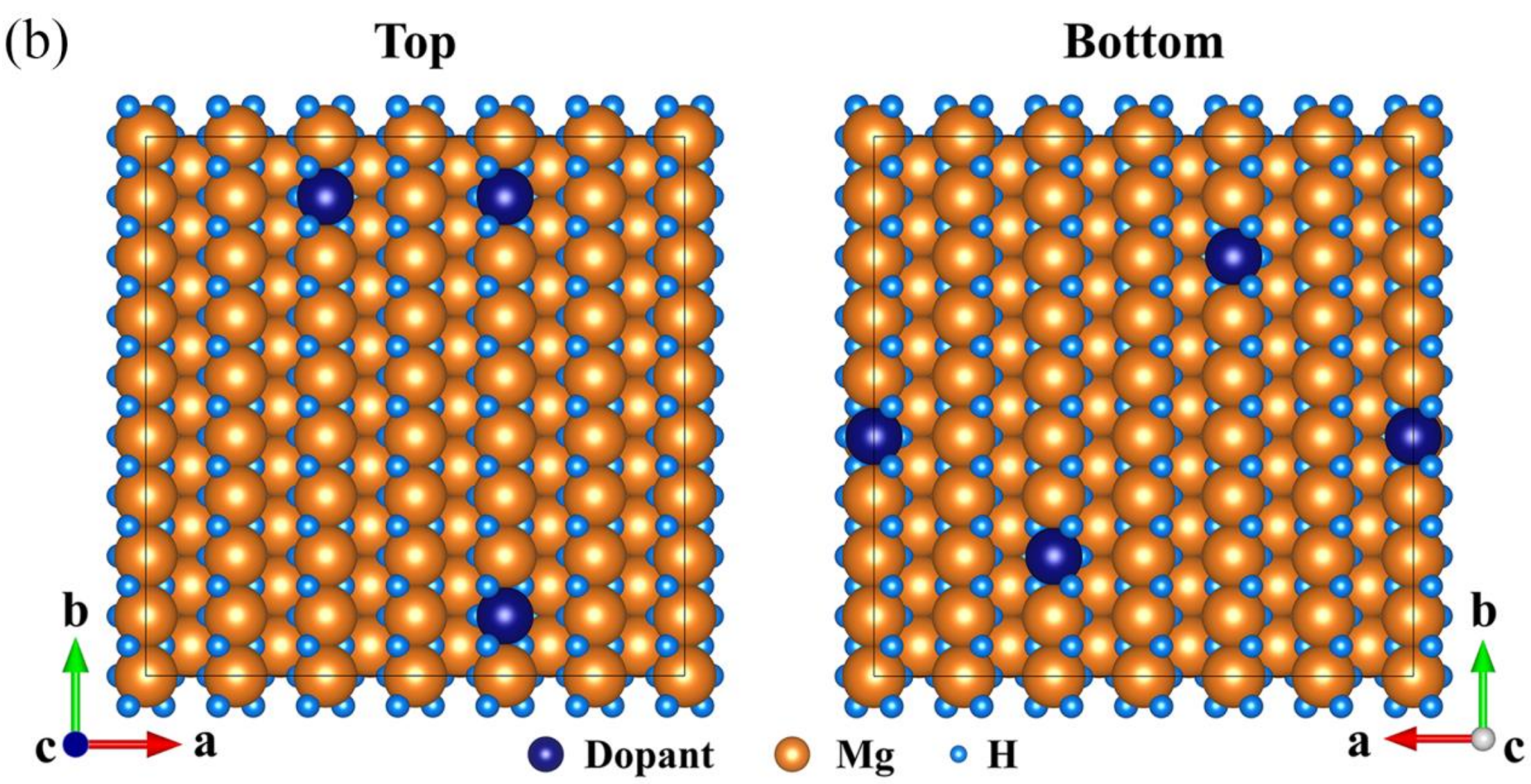

Fig. 4. (a) Selected doping elements for surface doping study; (b) Atomic structures of the doped $MgH_2(100)$ surface viewed from the top (left) and bottom (right). The large orange, small light blue, and large dark blue spheres represent Mg, H, and dopants, respectively.

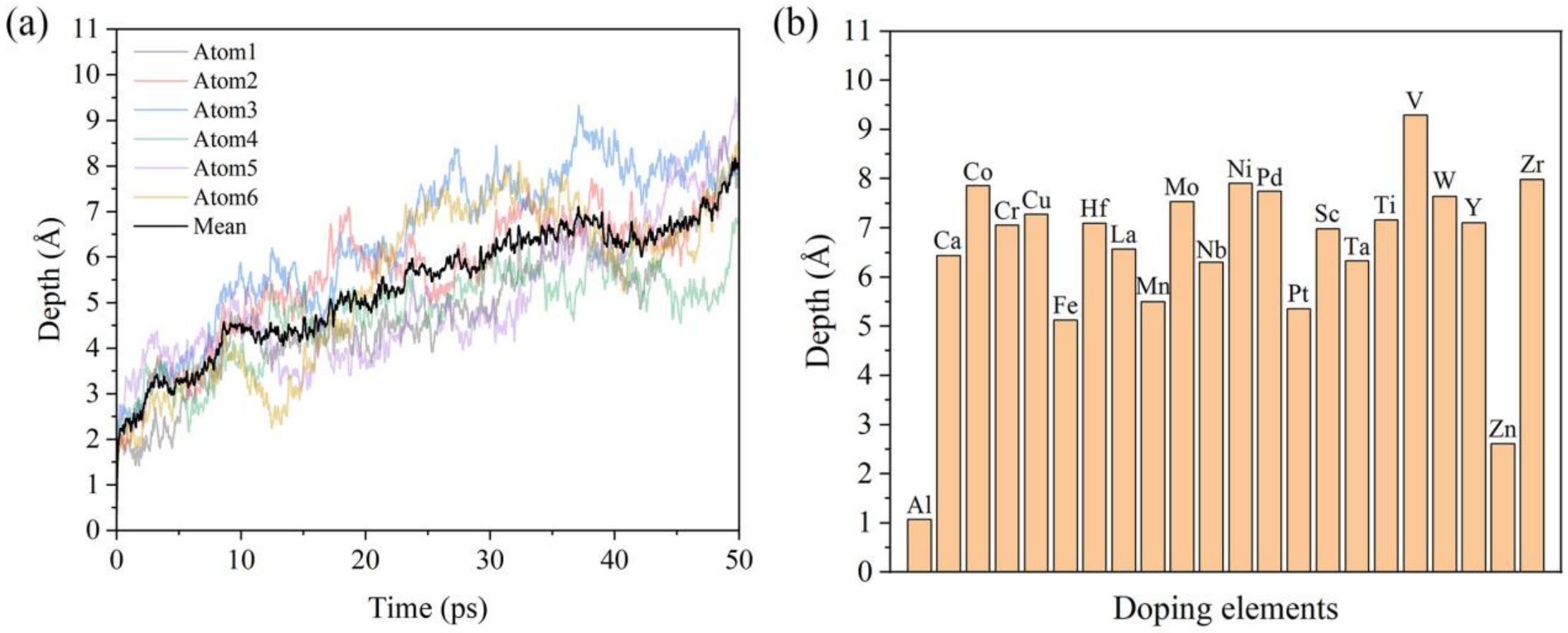

Fig. 5. Depth evolution of dopants during MD simulations. (a) Time-dependent mean depth of Ni dopants. (b) Mean depths of all screened doping elements at 50 ps.

The cumulative $H_2$ release profiles (Fig. 6, obtained by averaging several independent MD runs) reveal substantial variation in catalytic activity across different dopant species. Among all screened elements, Ni exhibits the highest catalytic activity, with both the greatest total hydrogen release and the fastest initial desorption rate. This finding is consistent with numerous experimental reports that have highlighted the superior catalytic effect of Ni on the dehydrogenation kinetics of $MgH_2$ [38-40]. Furthermore, theoretical studies have also corroborated the excellent enhancement effect of Ni on the dehydrogenation properties of $MgH_2$ [41, 42]. Quantification and discussion of the DSC-determined activation energies for desorption of Ni-containing versus pure Mg based hydride will be provided at the end of section 3.5.2 to further support this view. Experimentally, this exceptional activity is typically attributed to the in-situ formation of catalytic phases, specifically $Mg_2Ni$ and $Mg_2NiH_4$ [43-45]. These phases activate the so-called "hydrogen pump" mechanism. Specifically, these phases serve as an intermediary that continuously absorbs and releases hydrogen at the $MgH_2$/catalyst interface, effectively "pumping" hydrogen out of the $MgH_2$ matrix, which significantly accelerates the hydrogen diffusion and dissociation processes [46, 47]. Given this demonstrated efficiency, Ni continues to be a primary choice for the design of advanced composite catalysts [48-51].

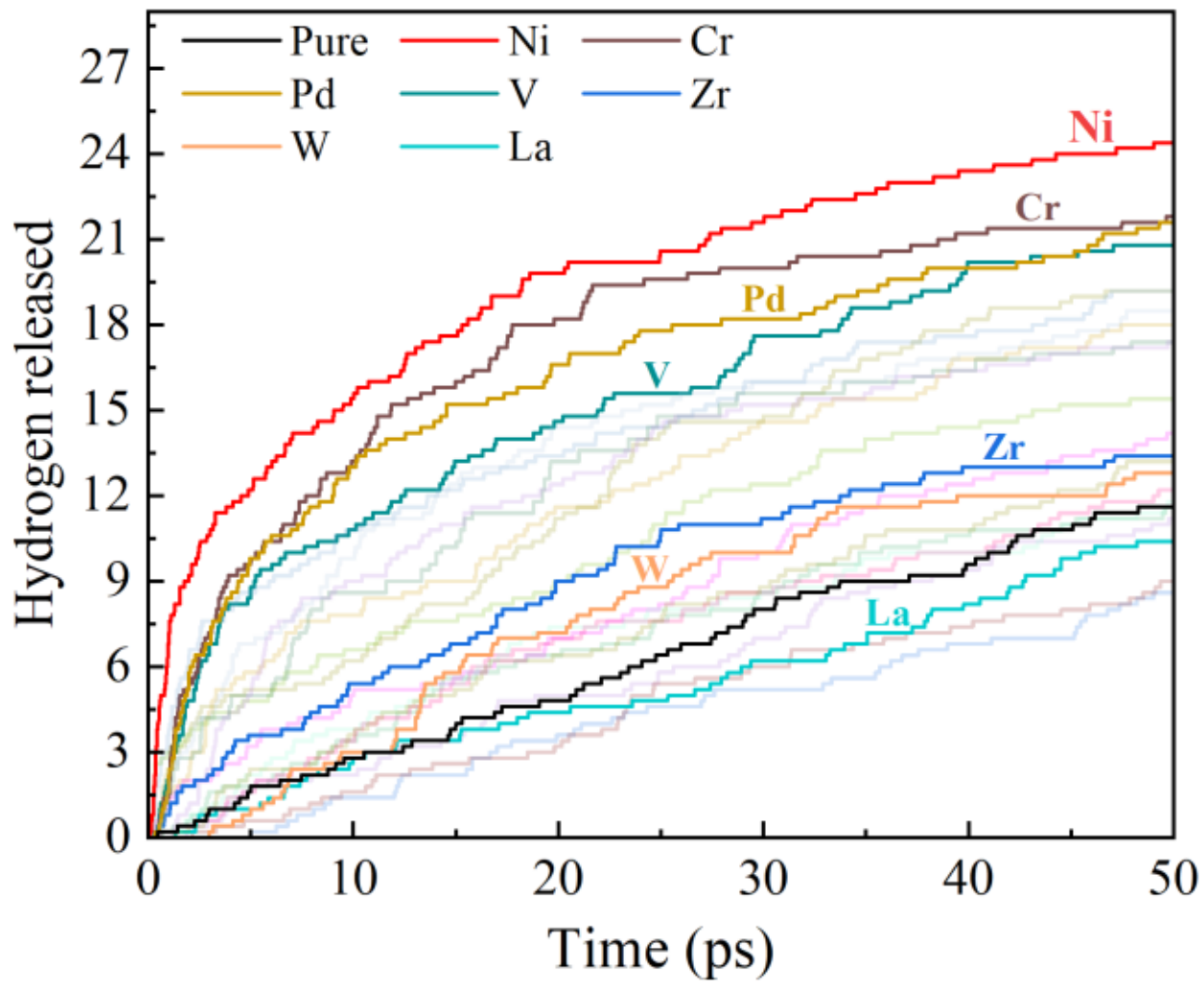

Fig. 6. The dehydrogenation curves of doped $MgH_2(100)$ surfaces obtained by averaging five independent MD runs for each doping element. For clarity, only selected curves (further discussed in the paper) are presented with full opacity and labels; other curves are shown with reduced opacity. A complete set of all dehydrogenation curves can be found in Fig. S3.

### 3.4 Model training and SHAP analysis

To reveal the atomic-scale origins of the observed activity trends, we performed machine learning-based descriptor analysis as described in Section 2.3. The trained XGBoost model achieved high predictive accuracy with a test $R^2 = 0.998$ (Fig. 7(a)), validating its capability to capture the relationships between dopant properties and dehydrogenation kinetics. The distribution of signed SHAP values for the considered atomic input features is shown in Fig. 7(b), while the mean absolute SHAP values summarized in Fig. S4 were used to quantify the overall feature importance. From both Fig. 7(b) and Fig. S4, the time-coupled Miedema electron density ($n_{ws}^{\tau}$) clearly emerges as the most influential descriptor governing the hydrogen release in doped systems. Its mean SHAP value (2.898) significantly exceeds those of the second and third most important descriptors: time-coupled electron affinity ($E_{ea}^{\tau}$, 1.329) and specific heat capacity ($C_p^{\tau}$, 0.830). This indicates that the effective electronic environment is critical for maximizing hydrogen desorption.

Next, to understand the leading role of $n_{ws}^{\tau}$, it is imperative to analyze its core constituent: the base Miedema electron density ($n_{ws}$). Therefore, the relationship between $n_{ws}$ of all doping elements and their respective total hydrogen release was plotted as shown in Fig. 7(c). The total hydrogen release refers to the mean cumulative number of $H_2$ molecules desorbed from the entire slab system (surface area: 752.4 $Å^2$) over 50 ps of averaged five MD simulations. This figure reveals that Miedema electron density exhibits a pronounced non-monotonic, volcano-shaped relationship to the total hydrogen release. Specifically, elements with lower $n_{ws}$ (e.g., La, Sc, and Zn) exhibit reduced total hydrogen release, while those with very high $n_{ws}$ (such as W) also show a decrease compared to elements within the optimal range (e.g., Ni, Cr, and Pd). This indicates that extreme values of $n_{ws}$, no matter whether high or low, are detrimental to dehydrogenation. Consequently, we identify an "optimal window" ($4.0 < n_{ws} < 5.4 \times 10^{-2}$ e/bohr$^3$) to effectively enhance the dehydrogenation kinetics of $MgH_2$. In stark contrast, electron affinity displays no such characteristic, showing instead an irregular relationship (Fig. 7(d)).

This analysis highlights that the observed non-monotonic dependence between Miedema electron density and dehydrogenation kinetics is fundamentally governed by the Sabatier principle [52, 53]. This principle dictates that optimal catalytic activity requires an intermediate strength of interaction, similarly as we have previously shown, e.g., for methane catalysis [54]. By enabling the identification and quantification of this non-linear interaction, where an optimal Miedema electron density provides the necessary balance for both effective H-H coupling and efficient $H_2$ desorption, our findings offer invaluable guidance. Indeed, our further calculations on the hydrogen release efficiencies of Ag- and N-doped $MgH_2$, explicitly marked in Fig. 7(c), strongly corroborate this finding, positioning these elements within the established volcano-type relationship. This principle-driven understanding is crucial for the rational design of materials with precisely tailored dehydrogenation kinetics, thereby effectively addressing complex, time-dependent dynamics in materials science applications. Furthermore, the Miedema electron density emerges as a robust descriptor not only for screening potential superior catalytic elements but also for guiding the rational design of more complex multi-component catalysts. In addition, the identified optimal $n_{ws}$ window appears consistent with observations from various hydrogen storage alloys. For instance, elements like Ni, Mn, and V, with $n_{ws}$ values falling within the above optimal window, are key constituents in some common hydrogen storage alloys like $LaNi_5$-based (AB5-type), $TiMn_2$-based (Laves phase), and V-based systems. Besides, the FeTi alloy (AB-type) serves as an example where this principle may apply: combining Ti (with low $n_{ws}$) and Fe (with high $n_{ws}$) may lead to a favorable mean Miedema electron density. This alignment suggests that adjusting the aggregate Miedema electron density by either selecting elements with intrinsically suitable $n_{ws}$ values or by appropriately combining complementary ones, can be a promising design principle for efficient hydrogen-storage alloys.

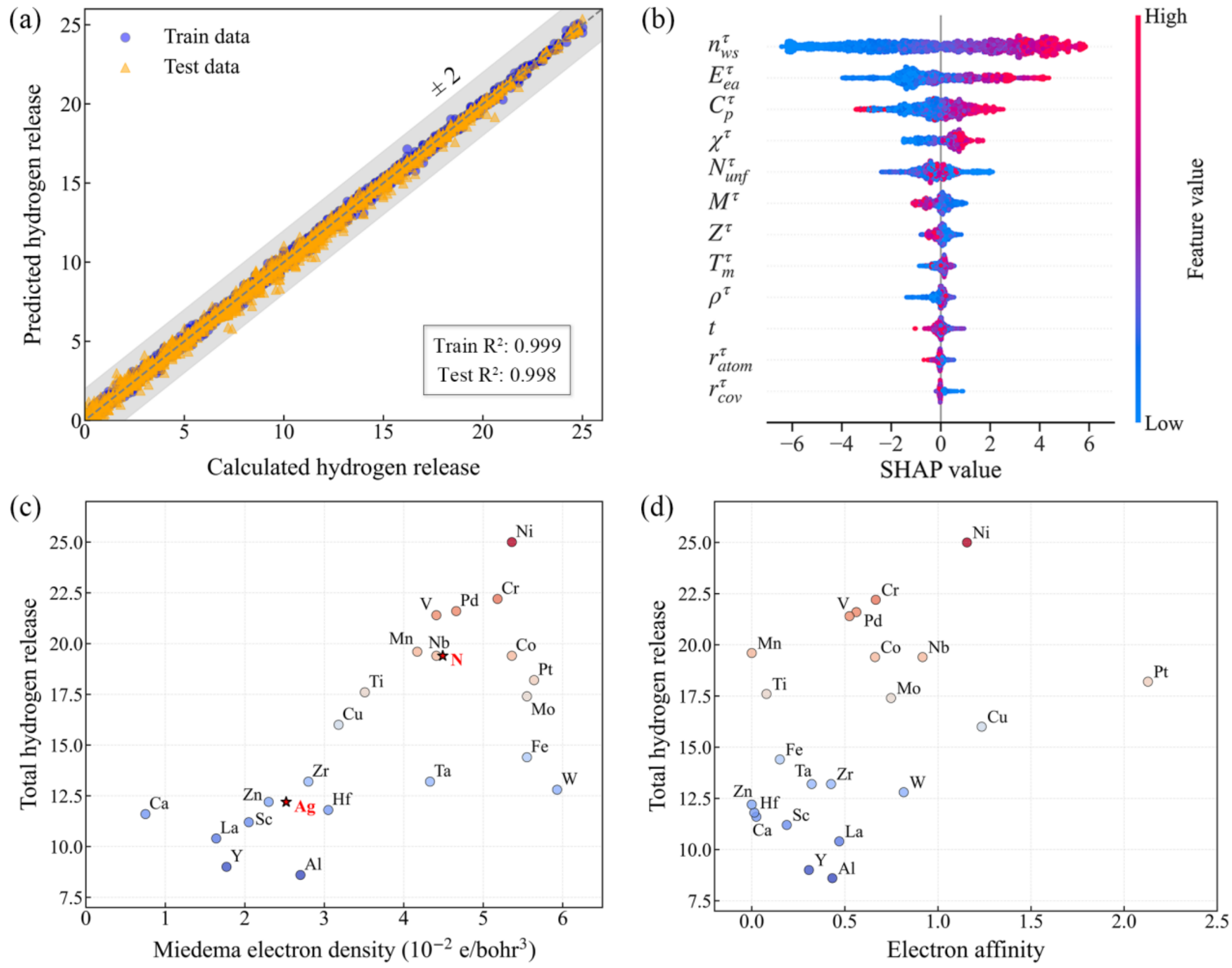

Fig. 7. (a) Comparison between calculated and predicted time-dependent hydrogen release values. (b) Distribution of signed SHAP values for the input descriptors. (c) Relationship between the Miedema electron density of the dopant and the total hydrogen release. (d) Relationship between the Miedema electron affinity of the dopant and the total hydrogen release. The color of the points in (c) and (d) represents the magnitude of hydrogen release, ranging from low (blue) to high (red).

### 3.5. Atomistic insight into the catalytic mechanism

#### 3.5.1 Spatial localization of $H_2$ formation

To further elucidate the atomic-level mechanism behind the enhanced dehydrogenation in doped $MgH_2$, and to connect it with the previously identified Miedema electron density ($n_{ws}$) trend, we conducted a detailed spatio-temporal analysis. We calculated the distance between each $H_2$ molecule and its nearest dopant at the moment of their formation, pooling data from five independent MD simulation runs. La, Zr, Ni, and W were selected as representative dopant species due to their distinct $n_{ws}$ values, which allowed for a systematic comparison regarding the hydrogen release trend illustrated in Fig. 7(c). The $n_{ws}$ values for La, Zr, Ni, and W are 1.64, 2.80, 5.36 and 5.93 × $10^{-2}$ e/bohr$^3$, respectively.

As evident from Fig. 8, an increasingly strong spatial correlation between the position of the dopant and the place of $H_2$ formation is observed as the Miedema electron density increases: at high $n_{ws}$, the vast majority of $H_2$ molecules, whether subsequently released (red) or not (black dots in the plots), are predominantly formed in the immediate vicinity of the dopants. This spatial localization clearly demonstrates that increasing $n_{ws}$ value leads to an increased catalytic activity for $H_2$ formation. At the same time, increased fraction of trapped (unreleased, black dost in the plots) $H_2$ molecules is evident with increasing $n_{ws}$. Dopants exhibiting favorable $n_{ws}$ values such as Ni possess specific electronic configurations that enable optimal interaction with H atoms, effectively attracting and binding them to their immediate vicinity, which facilitates effective $H_2$ formation (47 molecules per simulation). The balanced interaction then ensures the subsequent efficient $H_2$ release (25 molecules). Conversely, dopants characterized by unfavorable electron density struggle to establish the proper binding strength with H atoms for efficient attraction, thereby failing to provide both effective $H_2$ formation and subsequent release. This is compellingly demonstrated by the varied spatial distributions of newly formed $H_2$ molecules in the Zr- and W-doped systems. For the Zr-doped system, the weak interaction results in an insufficient attraction and aggregation of H atoms, leading to a marked lack of $H_2$ nucleation (21 molecules) and concentration around the dopant, but the $H_2$ molecules that do form are readily released (13 molecules). In contrast, for the W-doped system, the strong interaction indeed facilitates $H_2$ formation (82 molecules), but it concurrently prevents their efficient release (only 13 molecules). The strong correlation with the Miedema charge density and weak impact of the dopant size (Fig. 7c) indeed points towards dominant electronic impact rather than mechanistic effects related to the out diffusion of the $H_2$ molecules.

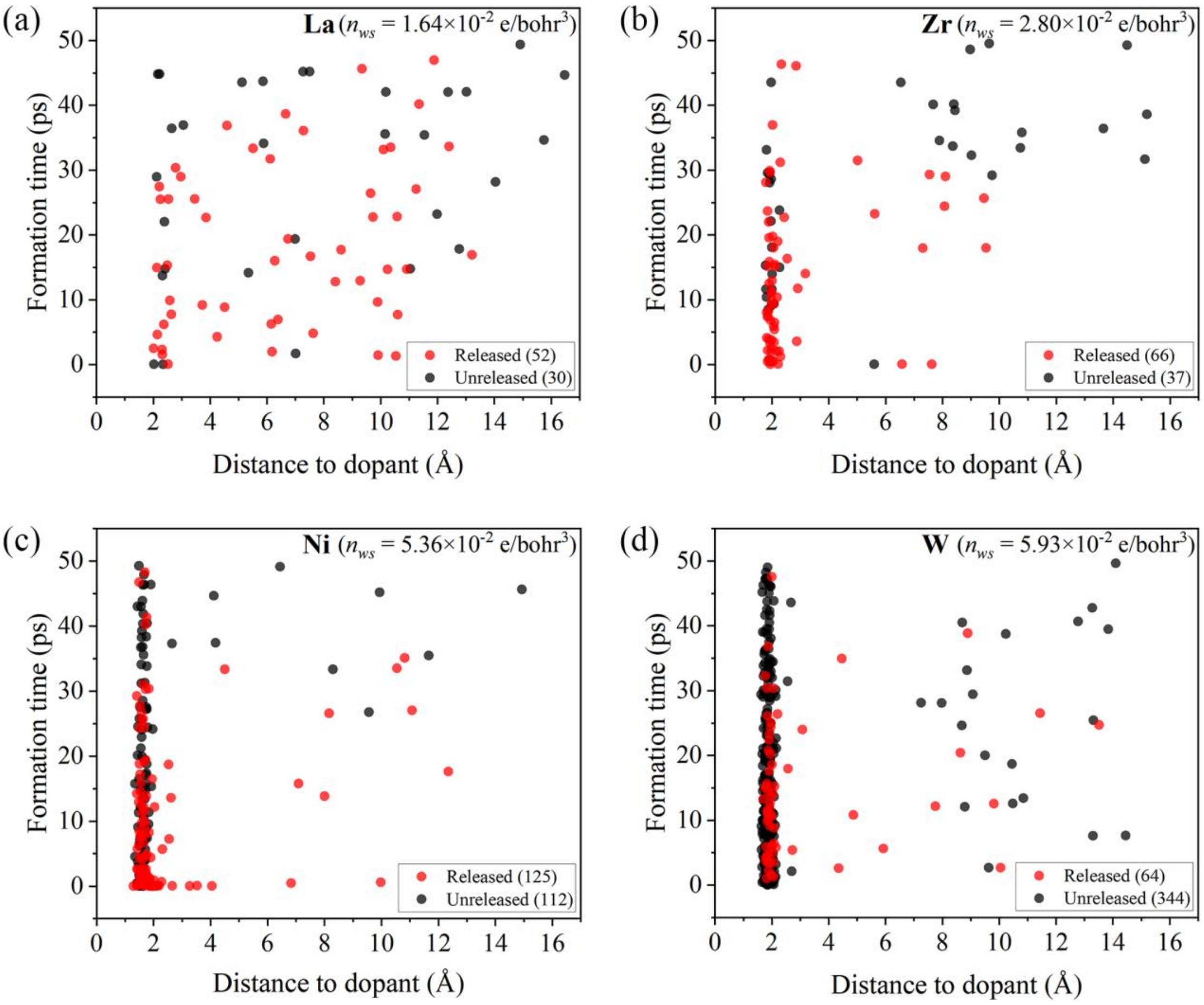

Fig. 8. Calculated distances between $H_2$ molecules at the moment of formation and their nearest dopants in $MgH_2$ systems doped with (a) La, (b) Zr, (c) Ni, and (d) W. Each subplot presents aggregated data obtained from five independent molecular dynamics (MD) simulations.

### 3.5.2 Hydrogen pump effect

Given the finite nature of the local H atoms in the proximity of dopants, the sustained $H_2$ formation observed in Fig. 8 necessitates the continuous replenishment of H atoms. To investigate this dynamical transport process, we performed radial distribution function (RDF) analysis of H atoms with respect to the Ni dopant. For each H atom in a given frame, we computed its minimum distance to any of the six dopants under periodic boundary conditions, and the resulting set of distances across the entire MD run was binned into a histogram with a bin width of 0.1 Å up to a cutoff of 10 Å. The RDF, $g(r)$, was then obtained by normalizing the histogram to the expected count for a uniform distribution at the same number density, such that $g(r)$ = 1 corresponds to a homogeneous H distribution and values greater than unity indicate local enrichment. To capture the temporal evolution, we divided the trajectory into three time segments and calculated the average RDFs—early (0-15%), middle (42-58%), and late (85-100%)—and averaged $g(r)$ over all frames within each segment. The final results represent the mean over five independent simulation runs.

As shown in Fig. 9(a), the RDF of Ni-H shows three distinct peaks throughout dehydrogenation. The peak positions remain essentially unchanged across the early, middle, and late stages, indicating a stable Ni-H bond length and a consistent local coordination geometry around the dopants. The first coordination peak is at ~1.57 Å, corresponding to H atoms in direct chemical contact with Ni atoms. Moreover, the first peak is exceptionally intense, indicating that H atoms are concentrated in the immediate vicinity of the dopants, which reflects a strong chemical affinity between dopants and hydrogen. This local enrichment establishes the dopants as thermodynamic sinks that attract and concentrate H atoms from the surrounding $MgH_2$ matrix, thereby providing the high local H density required for H-H recombination. This finding corresponds well to Fig. 8(c), which shows that the formed $H_2$ molecules are primarily located at distances of ~1.0-2.5 Å from the dopants. Additionally, the second and third coordination shell peaks show an increase in intensity in the middle and late stages compared to the early stage. This accumulation in the outer shells reveals the dynamical feeding process—H atoms migrate from the bulk $MgH_2$ matrix toward the dopant centers during dehydrogenation, building up a quasi-steady-state reservoir that continuously replenishes the first-shell H consumed by recombination reactions.

It is also worth noting that the blue and red curves representing middle and later stages, respectively, essentially overlap in Fig. 9. Our interpretation is that the dehydrogenation process becomes steady, i.e. equilibrated at the given conditions, at the middle time stage.

The combined picture that emerges is one of sustained directional transport: H atoms diffuse from the bulk towards the dopant sites, accumulate transiently in the second and third coordination shells, and are ultimately drawn into the first shell where they recombine to form $H_2$. The Ni atom thus functions simultaneously as a thermodynamic sink that attracts migrating H and as a kinetic facilitator that promotes H-H recombination, as evidenced by the preferential localization of newly formed $H_2$ molecules within the first coordination shell of the Ni dopant (Fig. 8(c)). Moreover, the dopant also facilitates subsequent $H_2$ release without causing product accumulation near the dopant site, as 125 out of 237 $H_2$ molecules (52.7%) formed across five independent MD runs successfully release (Fig. 8(c)). This dual role constitutes the microscopic basis of the experimental “hydrogen pump” mechanism [40, 41] and rationalizes the markedly enhanced dehydrogenation rates observed in the Ni-doped $MgH_2$ system.

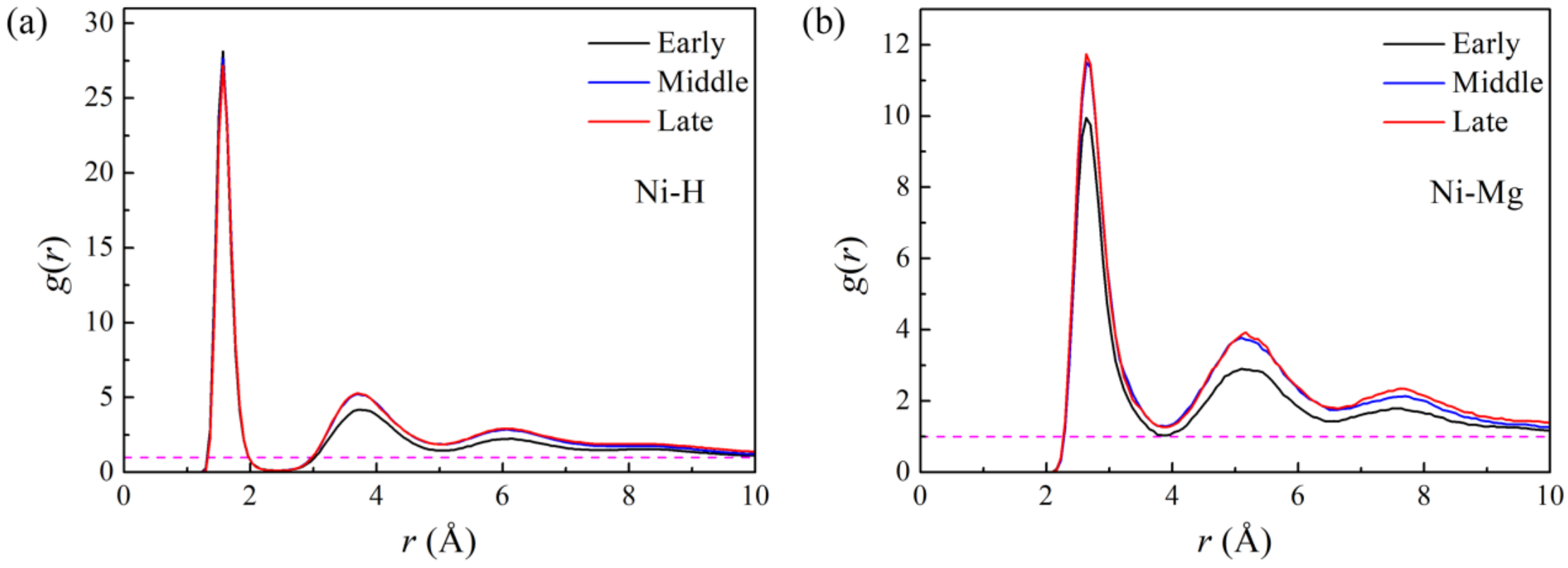

Fig. 9. Time-resolved radial distribution function $g(r)$ of (a) Ni-H and (b) Ni-Mg of Ni-doped $MgH_2(100)$ system, computed as the minimum H–dopant distance under periodic boundary conditions and normalized to the bulk H number density. Black, blue, and red curves correspond to the early (0-15%), middle (42-58%), and late (85-100%) stages of dehydrogenation, respectively, averaged over five independent trajectories. The dashed pink line marks the uniform distribution baseline ($g(r) = 1$).

Motivated by experimental observations that $Mg_2NiH_4$ serves as the catalytic phase in Ni-containing $MgH_2$ systems [43-45, 55], we further investigated whether similar local structures form in our simulations. To quantify the local coordination environment, the Ni-Mg pair RDFs were also computed using the same protocol as for Ni-H, as shown in Fig. 9(b). Besides, we calculated the coordination numbers of Ni-Mg and Ni-H by integrating the RDFs over their respective first coordination shells according to equation (2) [56]:

$$N_{\text{coord}} = 4\pi\rho \int_{r_{\min}}^{r_{\max}} r^2 g(r) \mathrm{d}r$$

(2)

where $\rho = N/V$ is the number density of the coordinating species ($N$ being the total number of Mg or H atoms in the system and $V$ the simulation box volume), and the integration limits [$r_{\min}$, $r_{\max}$] correspond to the first coordination shell as determined from the onset of the first peak to the first local minimum in $g(r)$ (2.03-3.90 Å for Ni-Mg, 1.16-2.43 Å for Ni-H).

From Fig. 9(a), Ni-H RDFs' first peaks locate at ~1.57 Å, in good agreement with the Ni-H nearest-neighbor distance in $Mg_2NiH_4$ (1.548 Å) [57, 58]. Similarly, Ni-Mg RDFs' first peaks appear at ~2.63 Å, in a reasonable agreement with the Ni–Mg nearest-neighbor distance in $Mg_2NiH_4$ (2.82 Å) [59]. For the Ni–Mg coordination, we obtain values of 6.14 (early), 7.18 (middle), and 7.22 (late), showing an increase from early to middle stages before stabilizing. Again, the rather small difference between middle and late stages suggests reaching steady process at those stages Combined with the Ni–H coordination numbers (early: 4.71, middle: 4.64, late: 4.55), the total Ni coordination environment at peak activity, composed of approximately 7-8 Mg plus 4-5 H atoms, closely resembles the local structure of $Mg_2NiH_4$ [56-58]. What is more, the temporal evolution of the coordination numbers reveals a dynamic process of forming the catalytically active $Mg_2NiH_4$-like cluster.

To experimentally verify the catalytic effect suggested by our simulations, we prepared pure $MgH_2$ and Ni-containing $MgH_2$ ($MgH_2$-14.9 wt.% Ni) samples and performed DSC measurements on them. As shown in Fig. 10(c) and Table 2, the pure sample exhibits a single desorption peak with an onset desorption temperature of 413 °C at a heating rate of 10 K/min and an activation energy of 171 ± 6 kJ/mol. In contrast, the Ni-doped sample shows a bimodal DSC signal with a reduced onset desorption temperature (349 °C) at a heating rate of 10 K/min. Deconvolution of the two peaks yields activation energies of 135 ± 3 kJ/mol and 101 ± 3 kJ/mol, both lower than that of pure $MgH_2$. It should be noted that the DSC profile of the Ni-containing $MgH_2$ sample recorded at a heating rate of 5 K/min exhibited only a single prominent peak. To prevent the introduction of mathematical artifacts from forced deconvolution and to ensure the high accuracy of the calculated activation energies, this specific dataset was excluded from the deconvolution process. Consequently, the Kissinger analysis for the bimodal peaks was performed using exclusively the true peak temperatures derived from the 2.5, 10, and 20 K/min measurements. The bimodal feature and reduced activation energies are consistent with the picture that Ni introduces (sub)stoichiometric, low-hydrogen-content entities (of the general type $Mg_2NiH_x$, $1 < x < 4$) that offer statistically distributed H interstitials, thereby enhancing diffusional hydrogen transport compared to the strongly localized ionic H positions in pristine $MgH_2$. The observation that only the Ni-containing material exhibits a superposition of desorption peaks (occasional bimodal DSC signals) further confirms the unique catalytic effect of Ni. These energetically similar but not completely congruent release processes can be understood as arising from the energetic fine structure of hydrogen interstitials in a heteroatom-doped system, where slightly different local environments give rise to distinct desorption barriers. This interpretation is in good agreement with our simulation results, where Ni dopants spontaneously form $Mg_2NiH_4$-like short-range clusters that promote $H_2$ formation and release (long-range $Mg_2NiH_4$ crystallites are not observed in our simulations due to the limited time and length scales, but may coexist in experiments).

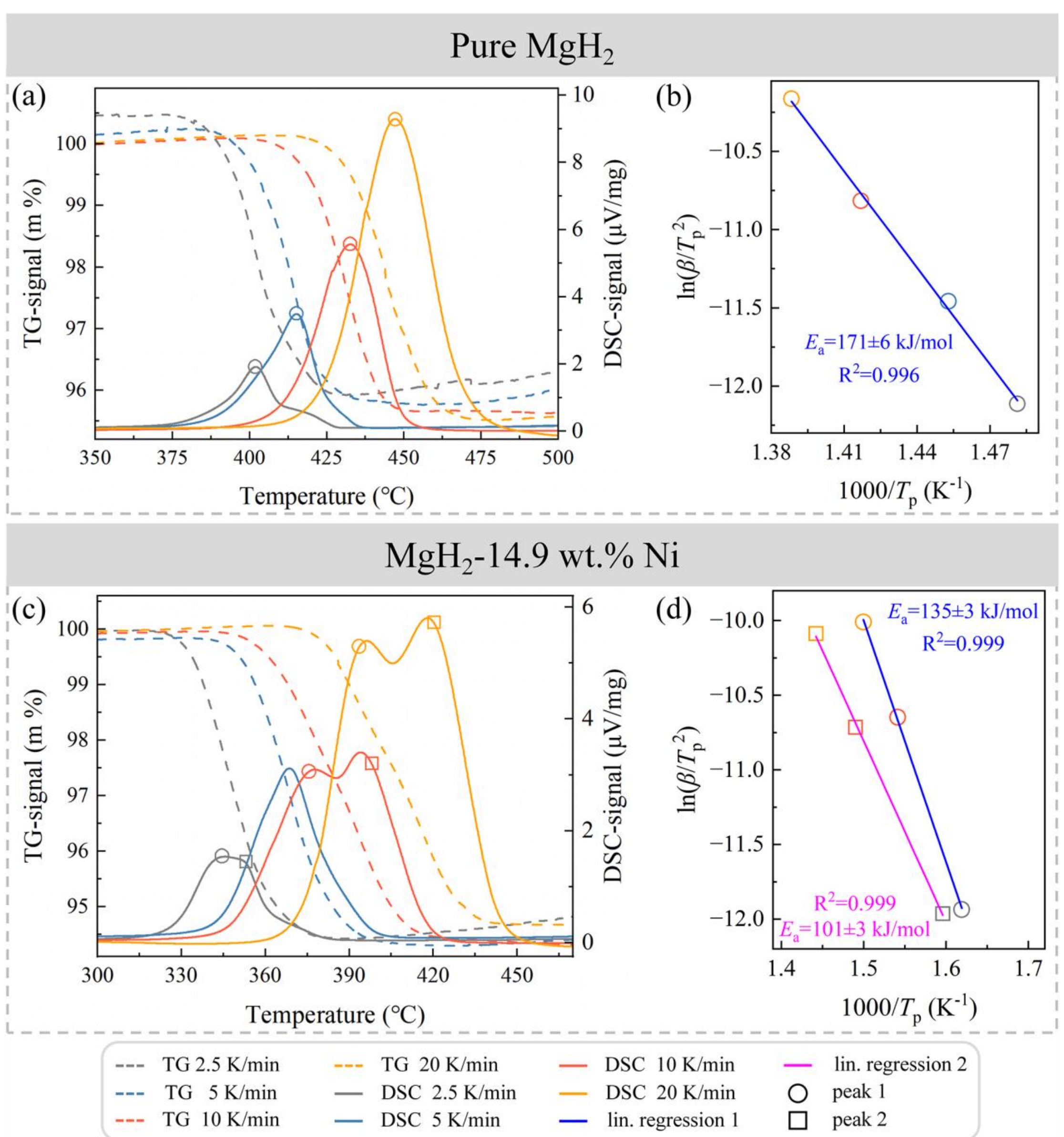

Fig. 10. Simultaneous thermal analysis of (a) pure $MgH_2$ and (c) $MgH_2$-14.9 wt.% Ni at various heating rates with marked peak temperatures. Corresponding Kissinger-plots (b) and (d) to estimate the desorption activation energy. Note that for $MgH_2$-14.9 wt.% Ni, two superimposed peaks result in a set of two desorption activation enthalpies after the peak deconvolution (see Fig. S5-S7).

## 4. Conclusions

In this study, we employed universal machine-learning potential molecular dynamics simulations to provide atomistic insights into $MgH_2$ dehydrogenation. Specifically, we systematically investigated the effect of surface orientation on the dehydrogenation of $MgH_2$ and its underlying mechanism. Furthermore, the impact of surface doping on dehydrogenation kinetics and catalytic mechanisms was comprehensively explored. From our calculations, the following key conclusions are drawn:

1. $MgH_2(100)$ is identified as the most active low-index surface for dehydrogenation. Additionally, investigation of the pristine $MgH_2(110)$ surface reveals a distinct $H_2$ formation mechanism, in which $H_2$ nucleation occurs in the subsurface region, followed by diffusion to the surface for desorption. This pathway significantly enriches the current understanding of $H_2$ evolution, highlighting the critical involvement of subsurface processes in addition to conventional surface-driven reactions.
2. Machine-learning analysis identifies Miedema electron density ($n_{ws}$) as the dominant physical descriptor governing catalytic activity. We established a quantitative volcano-shaped relationship between Miedema electron density and hydrogen release, defining an optimal window ($4.0 < n_{ws} < 5.4 \times 10^{-2}$ e/bohr$^3$). Spatial analysis of $H_2$ formation further revealed that dopants with optimal Miedema electron density serve a dual role: acting as a thermodynamic sink for hydrogen attraction and a kinetic facilitator for H-H coupling. This confirms the Sabatier principle active in this system: the dopant must exhibit sufficient interaction to efficiently attract H atoms and facilitate H-H coupling, yet avoid excessive binding that would hinder the subsequent desorption and release of the formed $H_2$ molecules. Our analysis identifies Ni as an optimum catalyst; the improved dehydrogenation kinetics in Ni-doped $MgH_2$ as compared with the pristine state has been confirmed experimentally.
3. Moreover, a detailed local structural evolution analysis demonstrates that the Ni dopants dynamically form local catalytic atomic clusters, which are Ni-containing Mg-H complexes structurally similar to known catalytic hydride phases $Mg_2NiH_4$. These atomic clusters act as active sites for a “hydrogen pump” mechanism, where H atoms migrate from the bulk, accumulate around the clusters, combine into $H_2$, and are subsequently released. This atomistic validation provides strong theoretical support for the experimentally observed "hydrogen pump" effect.

In summary, this study not only demonstrates the immense power of universal machine learning potentials in validating and deciphering complex atomic-scale catalytic mechanisms and establishing quantitative property-activity relationships for dopant design, but also provides a robust framework for the rational design of high-performance catalysts for improving the dehydrogenation properties of $MgH_2$ and other hydrogen storage materials. Moreover, the alignment of this criterion with classic hydrogen storage alloys like $LaNi_5$ and FeTi further hints at its broader applicability to hydrogen storage material design.

## CRediT authorship contribution statement

**Bo Han:** Investigation, Formal analysis, Conceptualization, Writing - original draft. **Jianchuan Wang:** Writing - review & editing, Supervision, Funding acquisition. **Rui Zhang:** Investigation, Formal analysis, Software. **Martin Matas:** Writing - review & editing, Funding acquisition. **Elias Vigl:** Investigation. **Jing Tang:** Writing - review & editing. **Yong Du:** Writing - review & editing, Resources, Software. **Christian Weiss:** Writing - review & editing. **David Holec:** Conceptualization, Formal analysis, Resources, Software, Writing - review & editing, Supervision.

## Data availability

The data are available from the corresponding author upon reasonable request.

## Declaration of competing interest

The authors declare that they have no known competing financial interests or personal relationships that could have appeared to influence the work reported in this paper.

## Acknowledgements

This research was financially supported by the National Natural Science Foundation of China (Grant No. U20A20237) and China Scholarship Council (No.

2024063770137). This work was also supported by the QM4ST project funded by PJAC (Grant No. CZ.02.01.01/00/22_008/0004572, call Excellent Research) and the COLOSSE project funded by the European Union (Grant Agreement No. 101158464).